\documentclass[aps,prb,twocolumn,floats,showpacs]{revtex4}
\usepackage{graphicx}
\usepackage{color}
\usepackage{bm}

\newcommand{\eexp}{\mbox{e}^}
\newcommand{\half}{\mbox{\small $\frac{1}{2}$}}
\begin{document}
\title{Interferometric resonance signatures of Majorana bound states  }
\author{  Anatoly Golub and Baruch Horovitz}
\affiliation{ Department of Physics, Ben-Gurion University, Beer Sheva 84105 Israel\\   }
\begin{abstract}
We calculate the current noise power spectrum in a nanoscopic interferometer consisting of a  Majorana bound state (MBS) and a  localized spin.  We show that for large voltage (though less than the superconducting gap) several strong resonance peaks appear at frequencies that depend on the Zeeman splitting of the localized spin and on its tunneling to the localized spin. We also evaluate the differential conductance and find the unitary limit peak $2e^2/h$ at zero voltage as well as peaks at voltages corresponding to the resonances. We propose that detection of the resonances and related peaks in the differential conductance provide a strong support for the presence of an MBS.
\end{abstract}
\maketitle
\section{ Introduction} In recent years the exotic Majorana bound state (MBS) is the focus of investigations in condensed matter physics. Different platforms for obtaining an MBS and variety of setups for experimental observation were suggested \cite{kitaev,fu,fu2,das,das2,oreg,alicea,nagaosa,aliceaR}. In particular a zero bias peak (ZBP) in the conductance was predicted \cite{been,tanaka,law,flensberg}. The leading candidate is  semiconductor quantum wire in proximity with an s-wave superconductor - a system  that generates a topological superconductor (TS) with two MBS's at its ends. A signature of an MBS in tunneling data
has been detected in normal metal - TS junctions \cite{kouw,das3,ando,yazdani}, though the evidence is not conclusive \cite{lee}.  An alternative  setup has been suggested \cite{ueda} for detecting an Aharonov -Bohm (AB) interference between an MBS and a quantum dot, predicting structure in the tunneling data. Furthermore, zero frequency shot noise has been studied \cite{demler,us2,blanter}.

A significant inspiration for our present work are ESR-STM (Electron Spin Resonance - Scanning Tunneling Microscopy) data \cite{man,man2} where the power spectrum of the STM current shows a signal at the Larmor frequency. This spin resonance phenomena under stationary conditions is significant both as a method for single spin detection as well as a theoretical challenge.
 Recently we showed \cite{caso,us} that in a nanoscopic interferometer the spin-orbit interactions allow for an interference of two tunneling paths resulting in a resonance effect.

 In the present work we consider a nanoscopic interferometer of one MBS and a quantum dot, the latter having a Zeeman splitting (see Fig.1). The system consists of a metal lead, a Zeeman split quantum dot and an MBS at the edge of a topological superconductor, all forming an interference loop.
We evaluate the current noise power as a function of frequency which  has a number of strong resonances that depend on the Larmor frequency and on the tunneling strength between the MBS and the quantum dot. In particular we find a resonance at a renormalized Larmor frequency. We note that spin orbit coupling is not essential for having these resonances since the MBS itself provides spin mixing. We also evaluate the current-voltage relation and find a ZBP in the conductance, independent of magnetic field, as well as side peaks that shift with magnetic field. We find that the ZBP has the unitary limit $2e^2/h$ for all parameters, while the side peaks also reach this unitary limit for weak tunneling.

 We note that a ZBP can also occur with a magnetic impurity producing a Shiba state \cite{lee3} or in a class D disordered superconductors \cite{altland} or with other types of disorder \cite{lee2}. In a magnetic field the ZBP of a Shiba state shows, in principle, a Zeeman splitting while the disordered case \cite{altland} is insensitive to the magnetic field. We propose that detection of the following unusual phenomena provides a strong support, possibly conclusive, for the presence of an MBS: (i) Larmor related resonances in the current noise, (ii) Unitary limit for the ZBP, and (iii) conduction peaks at voltages related to the resonances, that approach the unitary limit for weak tunneling.

\section{ The Hamiltonian} The Hamiltonian of our system consists of the normal metal lead part $H_L$, the quantum dot $H_d$ and the tunnel couplings $H_T$ parts. The geometry is defined in Fig. 1; $t_R,t_L,w$ define the tunnel couplings between the MBS and dot, between the dot and normal lead, and between the MBS and the lead, respectively. $N(0)$ is the density of states in the normal lead and the resonance widths turn out to be $\Gamma_{L,R}=2\pi N(0)t_{L,R}^2<<t_{L,R}$ and $\Gamma_w=2\pi N(0)w^2<<w $, consistent with Golden rule estimates.
The normal lead has a voltage bias $V$; we assume that $V$ is large compared to all the above energy scales in the system, including the Larmor frequency  (though $V$ is below the superconducting gap). We also assume that the MBS is well separated from other  MBSs, e.g.  at the other end of a TS wire, and therefore neglect the coupling between them.  We write
the Hamiltonian in spin ($s$ matrices) and Nambu (particle- hole space, $\tau$ matrices) as
\begin{eqnarray}\label{Ham}
  H_{d}& =& \half d^{\dagger}(\varepsilon s_0+ H s_z)\times\tau_z d\label{Hd}\\
  H_T &=&\half [(t_Lc^{\dagger}(0) \hat{u}  +t_R\gamma \bar{V}^{\dagger}s_0)\times \tau_z d+
  \nonumber\\
 && w c^{\dagger}(0)s_0\times\tau_z\bar{V}_{\varphi}\gamma] + h.c \nonumber\\
 \hat{ u}&=&\eexp{is_z\tau_z\psi}\eexp{is_y\theta/2}\eexp{is_z\tau_z\chi}\label{u}
\end{eqnarray}
where $s_0,\tau_0,s_i,\tau_i$ $(i=x,y,z)$ are unit and Pauli matrices, respectively, and $2H$ is the Larmor frequency, including the g factor. The Hamiltonian $H_L$ of the normal lead  has a standard form \cite {us}.
 The lead and dot electron operators are of the form $c=(c_{\uparrow},c_{\downarrow},c^{\dagger}_{\uparrow},c^{\dagger}_{\downarrow})^T$ and the Majorana fermion operator $\gamma$ comes with the spinor
$\bar{V}_{\varphi}=\half(e^{i\varphi},e^{i\varphi},
e^{-i\varphi}, e^{-i\varphi})^T$=$\half\hat{V}_{\varphi}$; $\bar{V}=\bar{V}_{\varphi=0}$, the phase $\varphi$ is an AB phase defined by threading the interferometer with a magnetic flux.
The average energy level of the dot  $\varepsilon$ is chosen for now as $\varepsilon=0$.
\begin{figure}
\begin{center}
\includegraphics [width=0.45 \textwidth ]{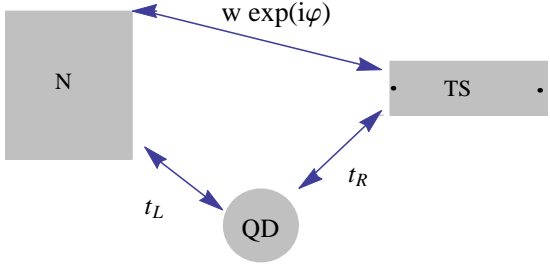}
\caption {Structure of the interferometer  which consists of normal metal lead, embedded quantum dot and topological superconductor. The  interaction couplings are presented. The AB  phase $\varphi$ accompanies the  direct tunneling  }
\end{center}
\end{figure}
 The spin orbit interaction matrix $\hat u$ corresponds to an SU(2) matrix $\hat u^{(2)}$ in spin space, which in the terms with $d^\dagger$ on the right becomes $\hat u^{(2)*}$, hence the $\tau_z$ factor in the exponents in (\ref{u}).
We include the spin-orbit matrix here for the sake of comparison with previous work \cite{us,caso}; in the present work we set ${\hat u}=1$ and comment on its possible effects below.

We note that in general the Majorana spinors $\bar{V}$ have a phase factor of the form $\exp[is_z\tau_z\nu]$, yet by redefining $c(0),d$ we can shift the phases $\nu$ into redefining the spin orbit phases $\psi,\chi$.

The current operator is defined as $J=e\frac{d}{dt}N_L=-ie[N_L,H]$ and acquires a form $J=(-i/4)(j_w +j_d)$ where
 \begin{eqnarray}
   j_w &=& w [c^{\dagger}(0)s_0\times\tau_0\bar {V}_{\varphi}\gamma-h.c.]\nonumber\\
   j_d &=& t_L[ c^{\dagger}(0)\hat{u}\times\tau_0 d-h.c.]
 \end{eqnarray}
 We use the current in Keldysh space \cite{keldysh,kamenev} $\hat{j}_w,\hat{j}_d $  to construct  the effective action  with source term. In the Keldysh theory the source field consists of two components: the classical $ \alpha_{cl}$ and quantum  one $\alpha$. The classical part $\alpha_{cl}$ is irrelevant for noise and current calculations and we put it to zero. In this case the source action  has a form
 \begin{equation}\label{source}
A
_{sour}=\frac{1}{4} \int_t \alpha (\hat{j}_w+\hat{j}_d)
\end{equation}
  After integrating out the lead and dot operators we arrive at the effective action in terms of Majorana Greens function (GF) which depends on coupling strengths and on quantum source  field $\alpha(t)$
 \begin{eqnarray}
   A_t &=& \frac{1}{2}\int_t \gamma^{T}G_{M}^{-1}\gamma; \,\,\,
   G_{M}^{-1} =G_{M0}^{-1}-\Sigma(\alpha)\nonumber\\
   \Sigma(\alpha)&=&\Sigma_1(\alpha)+\Sigma_2(\alpha)+ \Sigma_3(\alpha)+\Sigma_4(\alpha) \\
   \Sigma_1(\alpha) &=& \frac{\Gamma_w}{4} \hat{V_{\varphi}}^{\dagger}g_T \hat{V_{\varphi}};\,\,\, \Sigma_2(\alpha)=\frac{t_R^2}{4} \hat{V}^{\dagger}G_d \hat{V}\\
   \Sigma_3(\alpha) &=& \frac{\Gamma_w \Gamma_L}{4}\hat{V_{\varphi}}^{\dagger}g_TG_ u g_T \hat{V_{\varphi}}\\
    \Sigma_4(\alpha) &=&\frac{ w  \Gamma}{4}[\hat{V_{\varphi}}^{\dagger} g_T \hat{u} G_d \tau_z\hat{V}+
    \hat{V}^{\dagger}\tau_z G_d \hat{u}^{\dagger}g_T\hat{V_{\varphi}}]
 \end{eqnarray}
here $\Gamma=\sqrt{\Gamma_L\Gamma_R}$, $G_u(\alpha)=\hat{u}G_d\hat{u}^{\dagger}$, and $G^{R,A}_{M0}(E)=1/(E\pm i\delta)$.
The Keldysh  GFs of the lead is
\begin{eqnarray}
   g &= &\left(
                                                \begin{array}{cc}
                                                  g^R & g^K \\
                                                  0 & g^A \\
                                                \end{array}
                                              \right) ; \,\,\, g^{R,A} = \mp\frac{ i}{2}s_0\times\tau_0\\
 g^{K}&=&-i s_0 [\tanh(\frac{E+eV}{2T})\times P_+
 +\tanh(\frac{E-eV}{2T})\times P_-]\nonumber
\end{eqnarray}
($ P_{\pm}=(\tau_0\pm\tau_z)/2$) with the source contribution take the form  $g_T=T_- gT_+$, where
\begin{equation}\label{T}
    T_{\pm} = \tau_z\times\sigma_0\pm \alpha\tau_0\times\sigma_x/2
\end{equation}
 where matrices $\sigma_{x,y,z}$ are the Pauli matrices in Keldysh space.
 The GF of the quantum dot with quantum source  term is $G_d(E)=[G_{d0}^{-1}-\Gamma_L g_T]^{-1}$. If  $\alpha\rightarrow 0$ then the retarded component is
 \begin{equation}\label{Gd}
    G_{d}^R(E) = [(E+i\Gamma_L/2)s_0\times\tau_0- H s_z\times\tau_z]^{-1}
 \end{equation}

\section{ The currents noise power} We evaluate the current  and current noise spectral density by taking derivatives of the effective action with respect to $\alpha$. The complete derivation is presented in the appendices. Here we discuss the main results for the resonances in the current-current correlations $S(\omega)$.

 The noise power consists of two contributions $S=S_1+S_2$, Eqs. B1 in the appendix B. $S_1$ includes a single Majorana GF which has no structure (no spin) and  cannot show a spin resonance on its own. We note also that in a setup with only direct tunneling between the normal lead and Majorana state, i.e. $t_L=t_R=0$, the  $\omega$ dependence of the noise is weak. To see this let us take  the voltage  large (but below the superconducting gap) and temperature  $T\rightarrow 0$,  then the $S_2$ term  is exponentially small. The $S_1$ contribution is then a constant frequency independent noise:
\begin{equation}\label{S0}
   S_1=\frac{e^2\pi}{h}\Gamma_w
\end{equation}

  The significant part that is responsible for the resonance effect is $S_2$. This term  depends on two Majorana GF (see Eq. B2 in appendix B) and describes processes like those shown by the Feynman diagram in Fig. 2. This diagram belongs to a set whose hallmark are resonances related to a renormalized Larmor frequency. In the following we usually assume no spin orbit interaction i.e. $\hat u=1$ and equal tunnelings $t_L=t_R=w$. We also consider  the AB phase $\varphi=0$, legitimate for the magnetic fields used in the experiment \cite{kouw} $(H<0.14T)$ and the nanoscopic dimensions of our setup. Defining $p=(t_R/\Gamma)^2$ we find that to lowest order in the tunneling elements (i.e. $p\gg 1$) the contribution to the noise due to the process in Fig. 2 acquires the form
\begin{figure}
\begin{center}
\includegraphics [width=0.4\textwidth ]{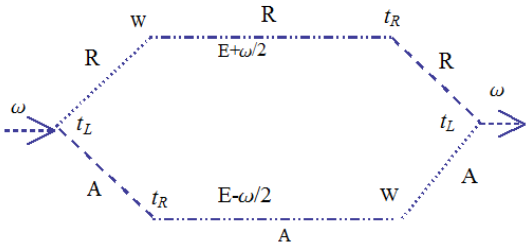}
\caption {Process that leads to resonances. Double dot - dashed lines stand for Majorana GFs, dashed lines represent quantum dot GFs and dotted lines denote normal lead GFs. Capital letter R and A correspond to retarded and advance functions, respectively.}
\end{center}
\end{figure}

\begin{eqnarray}
  P(\omega) &=& \frac{\pi p}{1+\omega^2}\frac{1+4H^2-3\omega^2}{(\omega^2-(2H+i)^2)(\omega^2-(2H-i)^2)}\nonumber\\\label{P}
\end{eqnarray}
We use dimensionless notations: all energies (including $H$) are taken in units of the tunneling width $\Gamma$ and the noise power is related to $P(\omega)$ by $S_2(\omega)=\frac{e^2}{h}\Gamma P(\omega)$ where $h$ is Planck's constant. We note the poles of the correlation function at the Larmor frequency $2H$ and in addition there are poles at $\omega=\pm i$ due to the MBS zero  energy state.

Returning now to the complete presentation of the correlation function we consider processes additional  to those in Fig. 2. These processes are determined by poles of both the dot and the Majorana GFs.  The later has a simple form (in dimensionless units)
  \begin{eqnarray}
  G_M^R(E) &=& \frac{(E+i/2-H)(E+i/2+H)}{(E+\frac{i\Gamma_w}{2\Gamma_L})((E+i/2)^2-\tilde{H}^2)+\xi}\label{GM}
\end{eqnarray}
where $ \tilde{H}=\sqrt{p+1/4 +H^2}$ is a renormalized magnetic field and $\xi=-is w H$ includes here spin orbit interaction, i.e. if the AB phase $\varphi=0$ then $s=\sin\theta/2\cos(\chi-\psi)$. For vanishing spin-orbit coupling $\theta=0$ we have $\xi=0$.

 We parameterize the renormalization of the Larmor frequency by $\lambda=\sqrt{p}/H=t_R/H$.
  As $\lambda$ increases the  poles at $\pm 2H$ in Eq.(\ref{P})   are renormalized. In particluar the Majorana GF has poles near $\pm \tilde H$ while its original pole at $E=0$ is maintained as a pole at $E=-i$. Other terms of the current spectral density  $S_2$ become more relevant as $\lambda$ increases.
E.g. the third term in Eq. B2 (appendix B)  which, unlike the one in Fig. 2,  consists of two Keldysh Majorana GFs (instead of retarded and advanced ones). This term is relatively small as $\sim\lambda^2$ if  $\lambda\ll 1$. We calculate $S_2$ considering all terms in Eq. B2 (appendix B) for various values of $\lambda$ and plot the corresponding $P(\omega)$ in Figs. 3 - 5; we use
 $H=250$ (in units of $\Gamma$).

We account for the position of the resonances in the following way: The poles for the dot GF in Eq. (\ref{Gd}) at $\pm H$ while those for the MBS Eq. (\ref{GM}) are at $0,\pm\tilde H$. The 5 resonance lines that we see in the power spectrum correspond to the following differences $\omega$ of these levels:
 \begin{eqnarray}\label{res}
 & 0\rightarrow 0 \qquad\qquad\qquad\qquad\qquad &\omega=0\nonumber\\
 & H\rightarrow \tilde H,\,-\tilde H\rightarrow -H  \qquad\qquad  &\omega=\tilde H-H\nonumber\\
 & 0\rightarrow H,\, -H\rightarrow 0\qquad\quad \qquad &\omega=H \nonumber\\
 & 0\rightarrow \tilde H,\, -\tilde H\rightarrow 0 \qquad \quad \qquad &\omega=\tilde H\nonumber\\
 & -H\rightarrow \tilde H,\,-\tilde H\rightarrow H\qquad\qquad &\omega=H+\tilde H
 \end{eqnarray}
 For small $\lambda$ these reduce to the Larmor frequency $2H$, half the Larmor frequency, and zero frequency. For finite $\lambda$ the Larmor frequency is renormalized to $H+\tilde H$ and corresponds to a transition from a dot level to a Majorana level. The peculiar "half Larmor" line consists of a negative signal at $H$ (transition between dot and Majorana states) and a positive signal at $\tilde H$ (transition between MBS states). We note that the lines at $2H,2\tilde H$ are missing.

\begin{figure}
\begin{center}
\includegraphics [width=0.45\textwidth ]{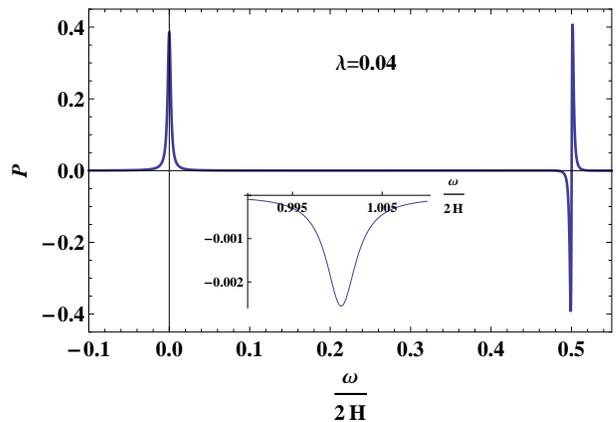}
\caption {Current noise for $\lambda$=0.04 and $H$=250 (equivalent to $p=10^2$). Note the peaks at half Larmor frequency and at $\omega=0$. The inset shows a small peak at the Larmor frequency. }
\end{center}
\end{figure}

\begin{figure}
\begin{center}
\includegraphics [width=0.45\textwidth ]{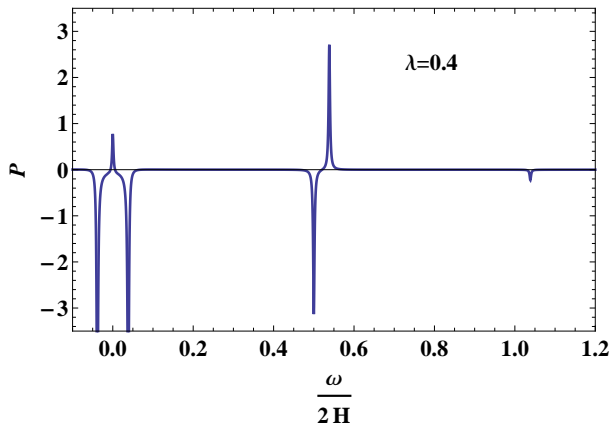}
\caption {The noise power  for $\lambda=0.4$ and $H=250$ (equivalent to $p=10^4$). The renormalized magnetic field $\tilde H=H\sqrt{1+\lambda^2}$ (for $H\gg\Gamma$) already affects the line positions. E.g. the last peak is at a renormalized Larmor frequency of $2.016 H$.}
\end{center}
\end{figure}

We note in Figs. 3 that the intensities of the lines at $\tilde H\pm H$ are small for small $\lambda$  and become visible on the scale of the other lines at $\lambda=0.4$ (Fig. 4). This conclusion is consistent with Eq. (\ref{GM}) and  the process presented in  Fig. 2. These lines involve $G_d^R(\pm H)G_M^R(\pm\tilde H)$ while the nominator of Eq. (\ref{GM}) involves the small $\tilde{H}-H$.
In contrast, all other transitions are strong even at small $\lambda$.
 We note that the positions of all resonances are well accounted by Eq. (\ref{res}) and do not depend on spin-orbit couplings since these frequencies correspond to resonance transitions within the dot-MBS system.

\begin{figure}
\begin{center}
\includegraphics [width=0.45\textwidth ]{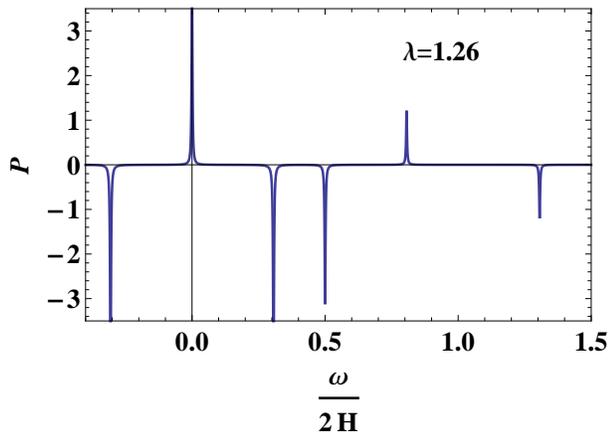}
\caption {The noise power  for $\lambda=1.26$ and $H=250$ (equivalent to $p=10^5$). Three resonances are   strongly shifted to higher frequencies.}
\end{center}
\end{figure}

We assumed equal tunnelings in the explicit calculation. For unequal tunnelings the intensities and linewidths are affected, yet the frequencies of the resonances are not affected, as in Eq.  (\ref{res}) ($\tilde H$ depends only on $t_R$). The resonances are due to an interference so that all tunneling $t$ that are either of $t_L,\,t_R,\,w$ must be finite and we expect that the intensities increase with either tunneling element. This can also be seen from Figs. 3-5 since $S(\omega)\sim\Gamma P(\omega)$ and increasing p implies decreasing $\Gamma$, i.e. $p\sim 1/t^2\sim 1/\Gamma$. We note, however, that the current noise $P(\omega)$, defined relative to the background Eq. (12), is actually increasing as tunneling is reduced, see Figs. 3-5. Hence an actual measurement would be more efficient with weak tunneling, e.g. as in Fig. 5 with $p=10^5$, i.e. $tN(0)\approx 10^{-3}$.

We note that for the case $\varepsilon\neq 0$ in Eq. (\ref{Hd}) where the quantum dot levels are shifted, the eigenvalues of the isolated Majorana and quantum dot system (a 5x5 matrix), i.e. $t_L=w=0$, can be easily solved. The differences between these levels yield the various resonances of the current noise, extending the result Eq. (\ref{res}). In particular the peak position at $\omega=0$ is independent of $\varepsilon$. Experimental detection of the resonance positions can determine the important parameters $\lambda$ and $\varepsilon$.

\section{ Conductance}
The dc current through the  interferometer (Fig. 1) for $\hat u=0$ (no spin-orbit coupling) and $\varphi$=0
acquires a simple form
 \begin{eqnarray}
    J&=&\frac{ie\Gamma_w}{8h}\int dE G_M^-(E)\Delta_-(E)[1+i\beta(E)]\label{J}\\
    \beta(E)&=&\frac{1}{2}(p-1/4)G_1^- (E)-\frac{(G_1^+(E))^2}{i-(p+1/4)G_1^-(E)}\nonumber
\end{eqnarray}
It is interesting to note that the  impact of the interferometer is  exhibited by function $\beta(E)$. The case $\beta(E)=0$ eliminates the QD (Fig. 1), a case that was considered earlier \cite{been,law,flensberg,us2}.
\begin{figure}
\begin{center}
\includegraphics [width=0.45\textwidth ]{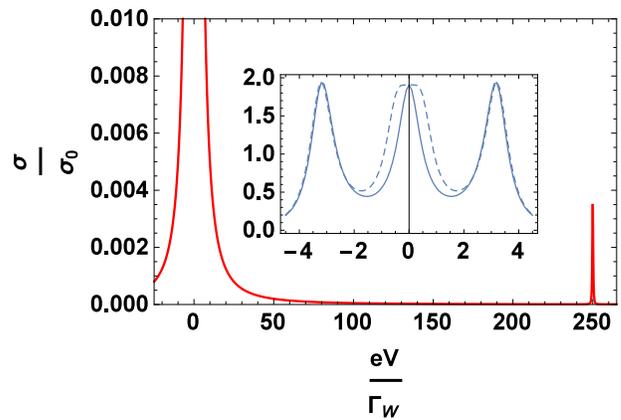}
\caption {The conductance $\sigma$ at $p=10$ and $T=0$ as function of voltage. External frame is for a magnetic field $H=250\Gamma$.  The ZBP reaches the unitary limit of $2\sigma_0=2e^2/h$, while weaker side bands appear at $\pm \tilde H,\pm H$ which overlap  (shown only for $V>0$). The inset shows $\sigma(V)$ for $H=0$ (full line) and $H=0.5\Gamma$ (dashed line).}
\end{center}
\end{figure}

\begin{figure}
\begin{center}
\includegraphics [width=0.45\textwidth ]{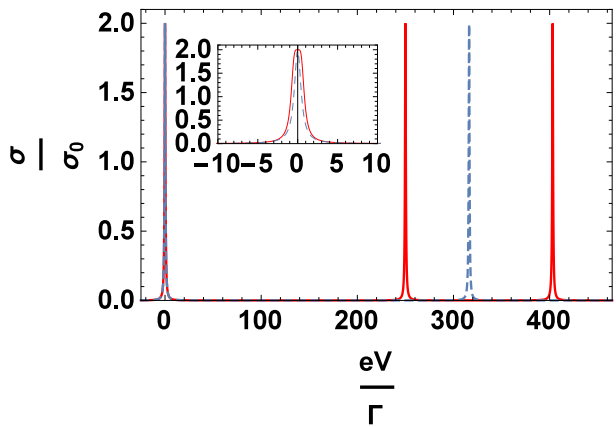}
\caption{ The conductance $\sigma$ at $p=10^5$ and $T=0$ as function of voltage for magnetic fields $H=0$ (dashed blue line) and $H=250\Gamma$ (full red line), the latter case resolving peaks at $H$ and $\tilde H$. Unitary limit is reached by all peaks for this $p=10^5$ (weak tunneling). The inset shows the ZBP for $H=0$ (dashed blue line) and $H=0.5\Gamma$ (full red line).}
\end{center}
\end{figure}
The conductance serves as additional probe for an MBS. In particular it shows a ZPB that reached the unitary limit $2e^2/h$ independent of the magnetic field or the tunneling value. We plot in Fig. 6 the conductance for $p=10$ (strong tunneling, close to parameters of Fig. 3). The inset with weak fields $H=0,0.5\Gamma$ shows peaks at $0,\,\pm\tilde H$ (peaks at $\pm H$ merge with the ZBP) while for $H=250\Gamma$ Fig. 6 shows peaks at $0,\pm H$ (the peaks at $\pm\tilde H$ now merge with those at $\pm H$); data is symmetric with respect to $V$. In the $H=250\Gamma$ case the side peak is much weaker than the ZBP. In Fig. 7 we plot the case $p=10^5$ (weak tunneling, as in Fig. 5) for $H=0$, showing two peaks at $0,\,\tilde H=\sqrt{p+1/4}$, and for $H=250\Gamma$, resolving 3 peaks at $0,H,\tilde H$. Remarkably, all peaks now reach the unitary limit.

We note that low temperature $T$ is essential for observing the ZBP, e.g. in Ref. \onlinecite{kouw} the ZBP is seen below $\sim 200$mK. The small side peak at eV=H for p=10 (Fig. 6) will be hard to see at such temperatures, yet, at weak tunneling $p=10^5$ the side peaks are strong and therefore as observable as is the ZPB. In contrast, the current noise is insensitive to temperature as long as it is below the voltage, which in turn is limited by the superconducting gap.

\section{ Conclusion} We have applied the standard Keldysh technique \cite{keldysh,kamenev} to evaluate the current noise spectral density in a nanoscopic interferomer consisting of a quantum dot, an MBS and a normal metal lead. We have found a number of resonance lines that uniquely  characterize the MBS. In particular there is a signal at a renormalized Larmor frequency, at a peculiar half renormalized Larmor frequency and at zero frequency.
The resonance lines at $\omega=0$ and near $\omega=H$ have an amplitude comparable to the background noise $S_1$ (Eq. \ref{S0}) even at small $\lambda$. At $\lambda\gtrsim 1$ (weak tunneling) all resonances are strong and comparable to $S_1$.

The inspiration for our setup is the ESR resonance measured in STM experiments \cite{man,man2}. In the latter case the spin-orbit interaction is an essential ingredient for generating an interference at the Larmor frequency \cite{caso,us}. In contrast, the MBS generates by itself spin mixing and a spin-orbit interaction is then not necessary for producing resonances in the current noise. The MBS manifests, in fact, the spin-orbit coupling characteristic of the topological superconductor that generates the MBS.  The presence of explicit spin-orbit coupling in our effective Hamiltonian Eq. (\ref{Ham}) may modify the intensities of the resonance lines, but not their positions, as given in Eq. (\ref{res}) for the $\varepsilon=0$ case. Evidently, control of the magnetic field and the dot-MBS coupling $t_R$ can provide a sensitive test for the MBS detection via our nanoscopic interferometer.

 We note that the MBS signature is due only to resonances that relate to the Larmor frequency, e.g. those in Eq. 15. Defects that are non-MBS may produce other types of resonances that are not of the Larmor type and therefore are irrelevant for MBS detection.
 A difficulty with experimental identification is due to the accuracy with which a zero energy state can be determined, as function e.g. of a magnetic field \cite{kouw,das3,ando,lee,yazdani}. Our interferometric method at large voltage allows for sharp line widths independent of temperature.
Therefore, control of the magnetic field and the dot-MBS coupling $t_R$  provide a sensitive  method for an MBS detection in our nanoscopic interferometer.

The main difficulty with experimental identification of an MBS via the method of a ZBP in conduction \cite{kouw,das3,ando,lee,yazdani} is that similar peaks may be due to other low energy bound states such as Shiba states \cite{lee3} or states localized by disorder \cite{altland,lee2}, or surface states as in d wave superconductors \cite{us3}. To support the presence of an MBS we propose detecting the following set of unusual phenomena: (i) Larmor related strong resonances in the current noise while the normal lead has zero or weak spin-orbit coupling, (ii) ZBP at unitary limit, independent of magnetic field (as is well known \cite{been,tanaka,law,flensberg}), and (iii) conductance peaks at voltages relating to the resonances in (i) that approach the unitary limit at weak tunneling. Zero bound states such as Shiba states \cite{lee3},  surface states as in d wave superconductors \cite{us3}, or the strong disorder case at low temperature \cite{lee2} do not satisfy criteria (ii). The ZBP of the weak disorder case \cite{altland} was shown to be insensitive to a magnetic field as it corresponds to interference in the spin singlet channel; hence it satisfies criteria (ii). We suggest that this case does not satisfy criteria (i) due to its singlet nature, however, we have not carried out an actual proof that requires evaluating our setup incorporating a class D disordered superconductor. We propose that all our three criteria are highly unusual and that their simultaneous detection is a strong support, probably conclusive, for presence of an MBS.

\begin{acknowledgments}
We would like to thank A. Altland, E. Demler, J. Sau, C. Kane and G. Zar\'{a}nd for stimulating discussions. This research was supported by THE ISRAEL SCIENCE FOUNDATION (BIKURA) (grant No. 1302/11) and the German-Israeli DIP project supported by the DFG.
\end{acknowledgments}

\appendix

\section{Green functions}

\label{append-dot}
 Here we calculate self energy functions in the effective action (see Eqs. (5-8) in the main text (MT)) which are the building blocks for Majorana GF .
  For retarded (advanced) parts  of $\Sigma^{R,A}$ we obtain
\begin{eqnarray}
 \Sigma_1^R &=& \frac{\Gamma_w}{4} \hat{V_{\varphi}}^{\dagger}g^R \hat{V_{\varphi}}=-\frac{i}{2}\Gamma_w \\
  \Sigma_2^R&=&\frac{t_R^2}{4} \hat{V}^{\dagger}G_d^R \hat{V}=t_R^2 G_1^R\\
  \Sigma_3^R &=& \frac{\Gamma_w \Gamma_L}{4}\hat{V_{\varphi}}^{\dagger}g^RG_u^R  g^R \hat{V_{\varphi}}\nonumber\\
  &&=\frac{\Gamma_w \Gamma_L}{8}[G_d^R(H)+G_d^R(-H)]\\
\Sigma_4^R &=& \frac{w  \Gamma}{4}[\hat{V_{\varphi}}^{\dagger} g^R u G_d^R \tau_z\hat{V}
+
    \hat{V}^{\dagger}\tau_z G_d^R u^{\dagger}g^R\hat{V_{\varphi}}]
 \end{eqnarray}
 and
 \begin{equation}
(\Sigma^R-\Sigma^A)_{1+2+3} =-i\Gamma_w+\frac{1}{2}(t_R^2
+ \frac{\Gamma_w \Gamma_L}{4})
(G_d^- (H)+G_d^- (-H))
\end{equation}
where the  quantum dot GFs are written as
 \begin{eqnarray}
G_d^{R,A}&=&G^{R,A}_1+G^{R,A}_2 s_z\tau_z; \,\,\,\, G_d^K =\Gamma_L G_d^R g^KG_d^A\nonumber \\
G^{R,A}_1&=&\frac{E\pm i\Gamma_L/2 }{(E\pm i\Gamma_L/2)^2-H^2}=\nonumber\\
&&\frac{1}{2}[G_d^{R,A}(H)+G_d^{R,A}(-H)]   \\
G^{R,A}_2&=&\frac{H}{(E\pm i\Gamma_L/2)^2-H^2}
\end{eqnarray}
and $G_d^{R,A}(H)=(E\pm i\Gamma_L/2 -H)^{-1}$. The interference part of self energy  $\Sigma_4^{R,A}$ is the only term of such type (R,A) that depend on spin-orbit interacting. By using the explicit formula for u-matrix we arrive at
\begin{eqnarray}
   \Sigma_4^{R,A} &=&\pm i w  \Gamma \{ G_2^{R,A} [\sin\frac{\theta}{2}\cos\varphi_- \cos\varphi \nonumber\\
   &&-\cos\frac{\theta}{2}\sin\varphi_+ \sin\varphi]\}
\end{eqnarray}
here $\varphi_{\pm}=\chi\pm \psi$. The Keldysh component of self-energy has spin-orbit interaction already in $\Sigma_3^K$
\begin{eqnarray}
  \Sigma_1^K &=&-\frac{i\Gamma_w }{2} \Delta_+ (E); \,\,\,\, \Sigma_2^K=
  \frac{t_R^2 }{4}\Delta_+ (E) G_1^-\\
\Sigma_3^K&=&\frac{\Gamma_w \Gamma_L}{4}[\Delta_+ (E)G_!^- \nonumber\\
&&-\sin(\theta)\cos2\psi \Delta_- (E) G_2^-]
\end{eqnarray}
The expression for $\Sigma_4^K$ is more involved
\begin{eqnarray}
 \Sigma_4^K &=& -i \frac{w  \Gamma}{2} [\Delta_- (E) G_1^+ \xi_1-\Delta_+(E)  G_2^+ \xi]\\
 \xi &=&-\cos\frac{\theta}{2}\sin\varphi_+ \sin\varphi+\sin\frac{\theta}{2}\cos\varphi_- \cos\varphi\nonumber\\
 \xi_1&=&\cos\frac{\theta}{2}\cos\varphi_+ \cos\varphi-\sin\frac{\theta}{2}\sin\varphi_- \sin\varphi
\end{eqnarray}

Here we use notations: $G_{d}^{\pm}=G_d^R \pm G_d^A$;  $G_{1,2}^{\pm}=G_{1,2}^R \pm G_{1,2}^A$. The Fermi factors are presented by functions $\Delta_{\pm}(E)=\tanh\frac{E+eV}{2T}\pm\tanh\frac{E-eV}{2T}$. In the absence of spin orbit interaction $\xi=0,\xi_1=\cos\varphi$.

Finally we come to the expression for retarded Majorana Gf presented in the MT by Eq. 13.

 The Keldysh component of Majorana GF acquires a form.
\begin{eqnarray}
  G_M^K (E) &=& \frac{G_M^-(E)}{2}\{\Delta_+ (E)+[\frac{\sin\theta \cos{2\psi}}{4}G_2^- (E)\nonumber\\
  &&+i\frac{w}{\Gamma_w} \xi_1 G_1^+(E) ]\frac{\Delta_- (E)}{\Sigma^R-\Sigma^A} \} \\
  \Sigma^R-\Sigma^A &=& -i+G_1^- (E) (p+1/4) +i\frac{w}{\Gamma_w}G_2^-(E)\xi\nonumber
\end{eqnarray}
where $p=(t_R/\Gamma_L)^2$.
\section{ Currents spectral density}
Noise power can be written as $S=S_1+S_2$
where \begin{eqnarray}
S_1 &=&\frac{e^2}{h} Tr[G_{MF}(t_1t_2)\frac{\delta^2\Sigma(t_2t_1)}{\delta\alpha(t)\delta\alpha(t')}]
\nonumber\\
&&=\frac{e^2}{h}(G_M^R\delta^2\Sigma_{11}+G_M^A\delta^2\Sigma_{22}+G_M^K\delta^2\Sigma_{21})\label{S1}\\
&&\nonumber\\
S_2 &=& \frac{e^2}{h}Tr[G_{M}(t_1t_2)\frac{\delta\Sigma(t_2t_3)}{\delta\alpha(t)}G_{M}(t_3t_4)
\frac{\delta\Sigma(t_4t_1)}{\delta\alpha(t')}]\nonumber
\end{eqnarray}
where Tr includes also the integration on time variables $t_1,t_2,t_3t_4$.
As we stressed in the MT our target  is  $S_2$ term which is responsible for resonances in the spectral density.
 Let  $\delta\Sigma=\varrho\Gamma_L/8 $, then taking trace in the  Keldysh space and performing the Fourier transform we obtain
\begin{eqnarray}
  S_2(\omega)&=&\frac{e^2\Gamma_L}{8h}\int dE\{G_{M}^R(E_-)\varrho^{11}(\omega)G_{M}^R(E_+)\varrho^{11}(-\omega)\nonumber\\
  &&+G_{M}^A(E_-)\varrho^{22}(\omega)G_{M}^A(E_+)\varrho^{22}(-\omega)+\nonumber\\
  &&\nonumber\\
&&G_{M}^K(E_-)\varrho^{21}(\omega)G_{M}^K(E_+)\varrho^{21}(-\omega)+\nonumber\\
&&\nonumber\\
&&[G_{M}^K(E_-)(\varrho^{21}(-\omega)G_{M}^R(E_+)\varrho^{11}(\omega)+\nonumber\\
&&\nonumber\\
&&\varrho^{22}(-\omega)G_{M}^A(E_+)\varrho^{21}(\omega))+\nonumber
\end{eqnarray}
\begin{equation}\label{SS2}
  G_{M}^R(E_-)\varrho^{12}(\omega)G_{M}^A(E_+)\varrho^{21}(-\omega) +(\omega\rightarrow -\omega)]\}
\end{equation}


where $E_{\pm}=E\pm\omega/2.$
We present vertex functions in the expression for $S_2$  in the  case of absence of spin-orbit interaction ($\varphi_+=0, \theta=0$). Also we consider  the temperature  $T\rightarrow 0$ and large voltage $V>>T$. The last two conditions inspire approximations $\Delta_+=0$  and  $\Delta_-=2$. Thus we obtain
\begin{eqnarray}
  \varrho^{11}(\omega) &=& -4ix(\omega) \cos\varphi-\frac{2w}{\Gamma_L}\sin\varphi [G_d^R(E_+)-G_d^A(E_-)]_s\nonumber\\
    x(\omega)&= &1+\frac{1}{2}(p + 1/4 +
   \frac{i(\omega + i)}{2})[G_d^R(E_+)G_d^A(E_-)]_s\nonumber\\
  \varrho^{21}(\omega)&=& 4 x(\omega)\sin\varphi-\frac{2wi}{\Gamma_L}\cos\varphi [G_d^R(E_+)+G_d^A(E_-)]_s\nonumber\\
  \varrho^{12}(\omega)&=&-\varrho^{21}(-\omega) +i \sin\varphi [G_d^+(E_+)-G_d^+(E_-)]_s
\end{eqnarray}
where $[F]_s$ denotes the sum $F(H)+F(-H)$, also $\varrho^{22}(\omega) = -\varrho^{11}(-\omega)$.

Let us consider  only the direct tunneling between the MBS and normal lead takes place (no quantum dot).
The exact Keldysh components of the vertices  in this case are $\varrho^{21}=\varrho^{12}=0$
and
\begin{equation}\label{ver}
\varrho^{11}(\omega)=-2i\Delta_-(E_-);,\,\,\,\varrho^{22}(\omega)=2i\Delta_-(E_+)
\end{equation}
Only first two terms survive in the Eq. (\ref{SS2})
\begin{eqnarray}\label{S20}
     S_2(\omega)&=&\frac{e^2\Gamma_L}{2h}\int dE\{G_{M}^R(E_-)G_{M}^R(E_+)\nonumber\\
 && +G_{M}^A(E_-)G_{M}^A(E_+)\}\Delta_-(E_+)\Delta_-(E_-)
\end{eqnarray}
 However, they are relevant only at small voltage, while  in the limit we consider here (large V ) these terms are exponentially small (integration involves only retarded (advanced) product).
Therefore, the noise is defined by $S_1$ contribution which immediately follows from Eqs. (\ref{S1}) and (\ref{ver})
\begin{eqnarray}\label{S10}
    S_1&=&\frac{ie^2\Gamma}{4h}\int dE(G_M^-(E_+)+G_M^-(E_-))[1-\nonumber\\
    &&\frac{1}{4}\Delta_+(E_+)\Delta_+(E_-)]
\end{eqnarray}
At large voltage the second term in the brackets vanishes and explicit integration  results in the formula Eq. 12 of the main text. In the case the interferometer based setup such types of contributions serve as a background noise maintaining the positivity of the current spectral density.

We finally note that the last term in Eq. (\ref{SS2}) describes processes like those presented by diagram on Fig. 2.
in the main text.  Majorana GFs separate two vertex blocks each of them includes the trace over spins.
Therefore, the product has terms consists of quantum dot GFs with opposite spins. To plot Figs. 3-5 we consider all  terms in Eq. (\ref{SS2}).

As a last remark we point out that the Majorana fermion
in our setup causes proximity induced superconductivity in the quantum dot as it does in other hybrid systems with a superconductor \cite {kouw2}; this effect is incorporated in our exact result. We demonstrate this effect for a simple case when all tunnel couplings vanish except for $t_R$. The anomalous (superconducting) component of the quantum dot GFs  can be easily seen in second order perturbation theory. Here we present the exact result using the effective action:
\begin{eqnarray}
  A_{dot} &=& \frac{1}{2}\int_t d^{\dagger}G_{QD}^{-1}d \nonumber\\
  G_{QD}&=& [G_{d0}^{-1}+(s_0+s_x)\times (\tau_x-\tau_0)t_R^2 G_{M0}/4]^{-1}\nonumber\\
\end{eqnarray}
where the anomalous GFs can be simply read. Indeed, taking the explicit forms for $G_{d0}$ and $G_{M0}$ (Eqs.(\ref{Gd}) with $\Gamma=0$)
we find for anomalous part of the GFs for either $p$ wave or $s$ wave coupling
\begin{eqnarray}
  G^R_{QDp}(E) &=& \frac{t_R^2 s_0 \tau_x}{4\tilde{E}(\tilde{E}^2 - H^2 - t_R^2)} \\
  G^R_{QDs}(E)  &=& \frac{t_R^2  [s_x \tau_x(E^2+H^2)-s_y\tau_y 2EH]}{4\tilde{E}(\tilde{E}^2 - H^2)(\tilde{E}^2 - H^2 - t_R^2)}\nonumber
\end{eqnarray}
where $\tilde{E}=E+i\delta$. These results are for $\bar V$ in Eq. (\ref{u}), choosing $\bar V_\varphi$ instead induces a phase $\varphi$ in these anomalous GFs.

\end{document}